\documentclass[sigconf]{acmart}
\usepackage{adjustbox}
\usepackage{graphics}
\usepackage{amsmath}
\DeclareMathOperator*{\argmax}{arg\,max}
\usepackage{enumitem}
\usepackage[ruled,linesnumbered,vlined]{algorithm2e}
\usepackage{algorithmicx}
\usepackage{algpseudocode}
\usepackage{mathtools}
\usepackage{tabularray}
\usepackage{booktabs}
\usepackage{array}

\usepackage{multirow}
\usepackage[normalem]{ulem}
\useunder{\uline}{\ul}{}

\AtBeginDocument{%
  \providecommand\BibTeX{{%
    \normalfont B\kern-0.5em{\scshape i\kern-0.25em b}\kern-0.8em\TeX}}}

\acmConference[RecSys'25]{RecSys'25: The 19th ACM Conference on Recommender Systems}{September 22nd-26th, 2025}{Prague, Czech Republic}
\acmBooktitle{RecSys'25: The 18th ACM International Conference on Web Search and Data Mining,
September 22nd-26th, 2025, Prague, Czech Republic}

\newcommand{\equalcontribnote}{Equal contributions.}

\copyrightyear{2025}
\acmYear{2025}
\setcopyright{acmlicensed}\acmConference[RecSys '25]{Proceedings of the Nineteenth ACM Conference on Recommender Systems}{September 22--26, 2025}{Prague, Czech Republic}
\acmBooktitle{Proceedings of the Nineteenth ACM Conference on Recommender Systems (RecSys '25), September 22--26, 2025, Prague, Czech Republic}
\acmDOI{10.1145/3705328.3748058}
\acmISBN{979-8-4007-1364-4/2025/09}

\begin{document}

\title{Non-parametric Graph Convolution for Re-ranking in Recommendation Systems}

\author{Zhongyu Ouyang}
\authornote{\equalcontribnote}
\email{zhongyu.ouyang.gr@dartmouth.edu}
\affiliation{%
  \institution{Dartmouth College}
  \city{Hanover}
  \state{New Hampshire}
  \country{USA}
}

\author{Mingxuan Ju}
\authornote{This work does not relate to his role at Snap Inc.}
\authornotemark[1]
\email{mju@snap.com}
\affiliation{
  \institution{Snap Inc.}
  \city{Bellevue}
  \state{Washington}
  \country{USA}
}
\affiliation{
  \institution{University of Notre Dame}
  \city{Notre Dame}
  \state{Indiana}
  \country{USA}
}

\author{Soroush Vosoughi}
\authornote{Corresponding author.}
\email{soroush.vosoughi@dartmouth.edu}
\affiliation{
  \institution{Dartmouth College}
  \city{Hanover}
  \state{New Hampshire}
  \country{USA}
}

\author{Yanfang Ye}
\authornotemark[3]
\email{yye7@nd.edu}
\affiliation{
  \institution{University of Notre Dame}
  \city{Notre Dame}
  \state{Indiana}
  \country{USA}
}

\begin{abstract}
Graph knowledge has been proven effective in enhancing item rankings in recommender systems (RecSys), particularly during the retrieval stage.
However, its application in the ranking stage, especially when richer contextual information in user-item interactions is available, remains underexplored. 
A major challenge lies in the substantial computational cost associated with repeatedly retrieving neighborhood information from billions of items stored in distributed systems. 
This resource-intensive requirement makes it difficult to scale graph-based methods in practical RecSys. 
To bridge this gap, we first demonstrate that incorporating graphs in the ranking stage improves ranking qualities.
Notably, while the improvement is evident, we show that the substantial computational overheads entailed by graphs are prohibitively expensive for real-world recommendations. 
In light of this, we propose a non-parametric strategy that utilizes graph convolution for re-ranking only during test time.
Our strategy circumvents the notorious computational overheads from graph convolution during training, and utilizes structural knowledge hidden in graphs on-the-fly during testing.
It can be used as a plug-and-play module and easily employed to enhance the ranking ability of various ranking layers of a real-world RecSys with significantly reduced computational overhead.
Through comprehensive experiments across four benchmark datasets with varying levels of sparsity, we demonstrate that our strategy yields noticeable improvements (i.e., \underline{\textbf{8.1}\%} on average) during testing time with little to no additional computational overheads (i.e., \underline{\textbf{0.5}\%} on average). 
Code: \url{https://github.com/zyouyang/RecSys2025_NonParamGC.git}

\end{abstract}


\maketitle

\section{Introduction}
Recommender systems (RecSys) are ubiquitous in online applications and have redefined user experiences in product recommendation on e-Commerce platforms~\cite{wang2021dcn, schafer1999recommender,ouyang2024symbolic}, personalized advertising~\cite{gomez2015netflix, van2013deep, ju2025learning,ouyang2025humanlikepreferenceprofilingsequential}, and friend recommendation on social media platforms~\cite{ma2008sorec, jamali2010matrix, fan2019graph, ju2025revisiting}.
A typical recommender system is composed of two key stages: \textit{retrieval} and \textit{ranking}. 
The \textit{retrieval} stage aims to efficiently filter a massive item pool, which often consists of billions of candidates, down to a manageable subset of relevant items using lightweight models or approximate matching techniques. 
Collaborative filtering (CF) is widely employed at this stage, relying on historical user-item interactions to recommend items based on similar user behavior patterns~\cite{rendle2012bpr, wang2022towards, koren2021advances, ju2024does}.
Traditional CF methods are predominantly based on matrix factorization~\cite{rendle2012bpr, wang2022towards}, where users and items are represented by learnable ID embeddings used to reconstruct user-item interactions.

In contrast, in the ranking stage, the retrieved candidates are ranked using more sophisticated models that incorporate rich contextual information, such as user features (e.g., demographic data), item features (e.g., product descriptions), and interaction features (e.g., transaction timestamps). 
These models capture the variability in user preferences across different contexts in which interactions occur. 
In industry, for efficiency purposes, ranking is typically simplified as the click-through rate (CTR) prediction problem~\cite{he2017neural, guo2017deepfm, wang2017deep}, where models estimate the probability of user-item interactions given contextual features. 
Although simplified as a binary classification task---predicting whether a high-quality interaction will occur---the predicted probability is commonly used as a ranking score to determine the item order.

Recently, graph neural networks (GNNs) have demonstrated competitive performance on graph-structured data and have inspired a growing body of research focused on modeling user-item interaction graphs commonly found in RecSys~\cite{wang2019neural,he2020lightgcn,yu2022graph,ouyang2024improve}.
By capturing high-order connectivity and propagating collaborative signals across the user-item graph, GNN-based methods have significantly improved recommendation performance, particularly in the retrieval stage. 
Notable examples include NGCF~\cite{wang2019neural}, LightGCN~\cite{he2020lightgcn}, and SimGCL~\cite{yu2022graph}, which effectively leverage graph-based embeddings to enhance candidate generation with performance increased by up to $\sim$40\%, compared with methods without graphs~\cite{wang2019neural}.
However, despite their success in \textit{retrieval}, the application of GNNs in the \textit{ranking} stage remains relatively underexplored. 
This is largely due to the \textit{ranking} stage's reliance on rich contextual features and the prohibitive computational overhead of incorporating graph-based neighborhood information into more complex predictive models. 

A recent effort, denoted as graph convolution machine (GCM), incorporates graphs in the ranking stage ~\cite{wu2022graph} by directly utilizing a vanilla GNN to conduct message passing over the user-item bipartite graph.
In the graph where GCM conducts message passing, each user/item node is attached with its corresponding features, and each edge is attached with its contextual features of the interaction. 
Although GCM effectively improves ranking quality, it entails several issues that hinder it from being applied to industries:

\vspace{0.1in}
\noindent
\textit{(ii) Firstly,} GCM is \textbf{prohibitively expensive to train} in distributed machine learning infrastructures that most real-world industrial applications utilize.
In real-world scenarios, billions of users and items co-exist, and their corresponding ID embeddings are usually distributed across multiple machines.
GCM requires repetitive message passing over the user-item bipartite graph during the model training.
Hence, training over a single pair of user and item entails multiple queries (quadratic to the training batch size) of their neighbors' embeddings, incurring tremendous overhead due to the limited communication bandwidth between distributed machines. 
This quadratic computational complexity can be further aggravated by the fact that such an expensive operation is repeatedly executed at each training step. 

\vspace{0.1in}
\noindent
\textit{(ii) Secondly}, applying GCM to existing context-aware RecSys \textbf{requires tremendous engineering efforts} to support large-scale graph machine learning.
Existing industrial pipelines mostly explore deep models (e.g., DCN~\cite{wang2017deep}) that take i.i.d. tabular data as input, which is intrinsically different from non-i.i.d. graph data. 
It would be more desirable and applicable for a strategy to effectively inject graph knowledge that can be easily extended to RecSys that are well-established in existing industrial pipelines. 
Therefore, to target these aforementioned challenges, we aim to answer: 
\begin{center}
\textbf{How can we efficiently yet effectively incorporate graphs in the \textit{ranking} stage of existing RecSys?}
\end{center}

To bridge the aforementioned research gap, we begin by validating the potential synergy between contextual features and graph-based knowledge between users and items.
We conduct a proof-of-concept experiment on a widely adopted model commonly employed as a ranking component in industrial RecSys.
Specifically, we substitute the vanilla user and item embedding tables with a simple graph-based encoder that performs message passing over the user-item bipartite graph.
This substitution yields noticeable improvements in ranking performance across four benchmark datasets.
However, it also introduces a significant increase in computational overhead due to repeated message passing during training. 
Moreover, although this approach is broadly applicable, it requires model re-training and infrastructural changes, which limit its practicality in real-world deployments.

To address the limitations of the naive substitution approach, we diverge from the conventional practice of incorporating graphs during training.
Instead, we propose a test-time augmentation strategy that utilizes graphs only once at test time to significantly improve models' ranking abilities.
As our strategy operates exclusively at test time, it can be seamlessly applied to a wide range of well-trained models with minimal additional computational overhead. 
Specifically, given a well-trained target model and a user-item query: 
\textit{(i)} Construct a graph-based  similarity matrix between users and items;
\textit{(ii)} Identify a set of retrieved candidate users and items based on similarity;
\textit{(ii)} Form an augmented set of user-item pairs using these candidates and query the target model to obtain their predicted interaction probabilities; 
\textit{(iii)} Aggregate the predictions through a simple weighted fusion for the final score. 
Despite its simplicity, our strategy consistently improves the ranking effectiveness of the target model, incurring only negligible inference-time costs.
Through this simple yet effective scheme, we conduct extensive experiments demonstrating that our strategy significantly enhances the ranking ability of the target model while introducing minimal additional overheads to the overall pipeline.

\section{Non-parametric Graph Convolution for Re-ranking}

\begin{figure*}[ht]
  \centering
  \includegraphics[width=0.9\textwidth]{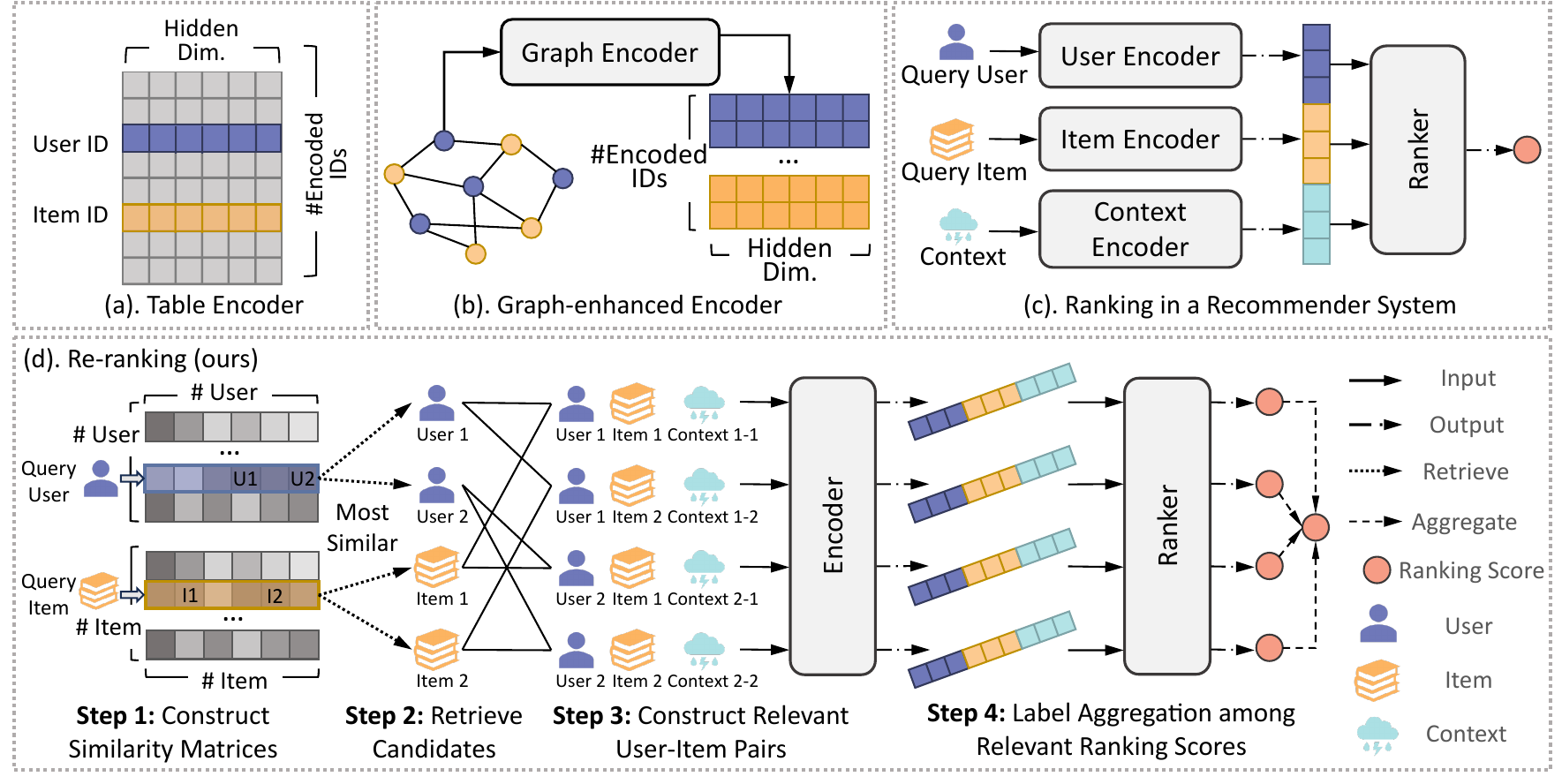}
  \caption{(a) The table encoder for user/item ID embeddings; (b) The graph-enhanced user/item ID encoder; (c) The calculation of ranking scores between users and items; (d) The overall framework of our strategy explained in steps.}
  \label{fig:model}
\end{figure*}

Without loss of generality, we adopt a CTR model as the ranking model, and illustrate the traditional paradigm of generating ranking scores without incorporating graph information in industrial RecSys.
Then, we present a graph-based encoder that integrates user-item connectivity into the models. 
Although improved ranking performance is observed, it incurs up to a $\sim$ 1000\% increase in the computational overhead in training, severely limiting its practicality. 
To overcome this issue, we introduce a test-time graph augmentation strategy that leverages graph knowledge solely during inference, effectively enhancing ranking quality without incurring additional training costs.
Our strategy is shown in Figure~\ref{fig:model}.

\subsection{Industrial Ranking Paradigms} \label{sec:cacf} 
Formally, we denote the IDs of user $i$ and item $j$ as $x_i$ and $x_j$ respectively, and the context features of the interaction between as $\textbf{c}_{ij} \in \mathbb{R}^{d^c}$, where $d^c$ refers to the dimension of the contextual feature.
Let the encoder of user/item IDs be $f(\cdot):\mathbb{R} \rightarrow \mathbb{R}^d$, where $d$ refers to the latent dimension of the encoded ID embeddings, and the contextual feature encoder be $h(\cdot): \mathbb{R}^{d^c} \rightarrow \mathbb{R}^{d'}$, where $d'$ is the dimension of encoded contextual embeddings.
When ranking with a CTR model, the problem is commonly simplified as a binary classification task, where we aim to predict the interaction probability ranging from 0 to 1.
We outline this paradigm in Figure~\ref{fig:model} (c).
To predict the probability of user $i$ and item $j$ having an interaction, the input $\textbf{z}_{ij} \in \mathbb{R}^{2d+d'}$ to a CTR model is defined as:
\begin{align}
    \textbf{z}_i & =  f(x_i), \; \textbf{z}_j = f(x_j), \; \textbf{z}^c_{ij} = h(\textbf{c}_{ij}), \label{eq:enc_1} \\
    \textbf{z}_{ij} & = \; \left[\textbf{z}_i \; \| \; \textbf{z}_j \; \| \; \textbf{z}^c_{ij}\right],
    \label{eq:enc_2}
\end{align}
where $\|$ refers to the concatenation operation, $\textbf{z}_i$ and $\textbf{z}_j$ refer to the latent ID embeddings of user $i$ and item $j$ generated by the encoder respectively, and $\textbf{z}^c_{ij}$ is the encoded contextual features. 
We depict this table embedding mechanism in Figure~\ref{fig:model} (a).
The three components of the input are then concatenated as the final input to the model, as depicted in Figure~\ref{fig:model}(c).

The traditional encoder $f(\cdot)$ in models in the ranking stage is a look-up table encoder, where $f(x_i)$ is the $x_i$-th entry in a matrix $E \in \mathbb{R}^{(|\mathcal{U}| + |\mathcal{I}|) \times d}$, $\mathcal{U}$ is the user set and $\mathcal{I}$ is the item set.
We depict the table encoder in Figure~\ref{fig:model}(a).
With the encoded embeddings, each CTR model then adopts a corresponding rating function $r(\cdot): \mathbb{R}^{2d+d'} \rightarrow \mathbb{R}$ to predict the interaction likelihood.
Specifically, we let $p_{ij} = r(\textbf{z}_{ij})$, where $p_{ij}$ represents how likely user $i$ would click/interact with item $j$.

To showcase the learning scheme of a ranking model taking contextual information into consideration, we demonstrate the modeling process of DCN~\cite{wang2021dcn}, one of the most typical CTR models utilized in industries for ranking.
Specifically, DCN comprises two main components: the cross network and the deep network.
The cross network, defined by its cross layers, is designed to model explicit feature interactions. 
Each cross layer is formulated as:
\begin{equation}
    \textbf{z}^{(l+1)} = \textbf{z}^{(0)}{\textbf{z}^{(l)}}^\intercal \textbf{w}^{(l)} + \textbf{b}^{(l)} + \textbf{z}^{(l)},
\end{equation}
where $\textbf{z}^{(l+1)}, \textbf{z}^{(l)} \in \mathbb{R}^d$ are the input and output column vectors from the $l$-th cross layer, and $\textbf{w}^{(l)}, \textbf{b}^{(l)} \in \mathbb{R}^d$ are the trainable parameters in the $l$-th cross layer.
On the other hand, the deep network is a series of fully connected layers designed to capture complex and non-linear interactions between the input features:
\begin{equation}
    \textbf{h}^{(l+1)} = \text{ReLU}(\textbf{W}^{(l+1)}\textbf{h}^{(l)} + \textbf{b}^{(l)}),
\end{equation}
where $\textbf{h}^{(l)} \in \mathbb{R}^{d_l}, \textbf{h}^{(l+1)} \in \mathbb{R}^{d_{l+1}}$ are the input and output of the $l$-th hidden layer, respectively, and $\textbf{W}^{(l)} \in \mathbb{R}^{d_{l+1} \times d_l}$ are the learnable parameters in the $l$-th layer.

For a specific user $i$ and item $j$, $\textbf{z}^{(0)} = \textbf{h}^{(0)} = \textbf{z}_{ij}$.
The outputs of both networks are then concatenated and fed to a two-class logits layer to generate the final prediction:
\begin{equation}
\label{eq:dcn-logits}
    p_{ij} = \sigma \left( [\textbf{z}^{(L_1)} \; \| \; \textbf{h}^{(L_2)} ] \textbf{w}_{\text{logits}} \right),
\end{equation}
where $\textbf{z}^{(L_1)} \in \mathbb{R}^{d_1}, \textbf{h}^{(L_2)} \in \mathbb{R}^{d_2}$ are the $L_1$-th and $L_2$-th layer outputs from the cross and deep network respectively, $\textbf{w}_{\text{logits}} \in \mathbb{R}^{d_1 + d_2}$ is the weight vector in the logits layer, and $\sigma(x) = 1 / (1 + e^{-x})$.
The DCN model is then trained with the binary cross entropy loss:
\begin{equation}
\label{eq:dcn-loss}
    \mathcal{L} = -\frac{1}{N} \sum_{\{i,j\}\in T_r} y_{ij} \log(p_{ij}) + (1-y_{ij}) \log(1-p_{ij}) + \lambda \sum_l \Vert\textbf{w}^{(l)} \Vert ^ 2,
\end{equation}
where $T_r$ denotes the training set of positively and negatively user-item interaction pairs, $p_{ij}$ is the predicted interaction probability between user $i$ and item $j$, $y_{ij}$ refer to the binary label (1 for positive and 0 for negative pairs), and $\lambda$ is the $L_2$ regularization coefficient.

\subsection{A Naive Graph-based Ranking Framework}  \label{sec:graph-enc}
In this section, we first introduce the definition of the user-item interaction bipartite graph that depicts rich topological relationships such as co-purchase and shared interests.
Then, we present how graph knowledge can be naively incorporated into existing ranking methods in their training processes.

Formally, let the interaction matrix be $\mathbf{M} \in \{0, 1\}^{|\mathcal{U}| \times |\mathcal{I}|}$, where $m_{ij}=1$ represents an observed positive interaction between user $i$ and item $j$, and $m_{ij}=0$ otherwise.
The interaction graph is defined as $\mathcal{G} = (\mathcal{V}, \mathcal{E})$, where $\mathcal{V} = \mathcal{U} \cup \mathcal{I}$ is the set of nodes, and $\mathcal{E}=\{(i, j) | \forall i \in \mathcal{U}, \forall j \in \mathcal{I}, m_{ij}=1\}$ is the set of edges.

A graph-based encoder utilizes the interaction bipartite graph to enhance the ID embedding quality. 
Unlike a table encoder which independently encodes the users and items, when generating the user and item embeddings, a graph encoder additionally leverages their graph relationships (e.g., co-purchase, shared interests, etc).
We depict the graph encoder in Figure~\ref{fig:model} (b).

Most existing graph-enhanced CF methods in the retrieval stage focus on scenarios without the incorporation of contextual features, such as NGCF~\cite{wang2019neural} and LightGCN~\cite{he2020lightgcn}.
However, whether or not their effectiveness can be transferred to existing ranking methods needs further investigation. 
Since the message passing mechanism~\cite{wang2019neural,he2020lightgcn,kipf2017semi} in graph-enhanced CF methods is the key to extracting graph knowledge, a natural way to extend ranking methods with graph knowledge is to include this mechanism in their paradigms.
Following this path, we adapt a well-studied linear message passing mechanism~\cite{he2020lightgcn} to existing ranking methods.

Specifically, let $f_g(\cdot, \cdot): \mathcal{G} \times x \rightarrow \mathbb{R}^d$ be the graph encoder.
For user $i$ and item $j$, $f_g(\cdot, \cdot)$ conducts message passing in each layer to propagate and aggregate information from the neighborhood. 
The graph-encoded embedding for node $i$ (user or item) is formulated as follows:
\begin{align}
    & \textbf{z}_i =  f_g(\mathcal{G}, x_i) =  \sum_{l=0}^{L}a_l \textbf{z}_i^{(l)}, \\
     \text{where} \;\;\; \textbf{z}_i^{(l)} = & \sum_{v\in N_i}  \frac{1}{\sqrt{|N_i|}\sqrt{|N_j|}} \textbf{z}_j^{(l-1)} \;\; \text{and} \;\; \textbf{z}_j^{(0)} = f(x_j),
     \label{eq:lightgcn}
\end{align}
In Equation~(\ref{eq:lightgcn}), $\textbf{z}_i^{(l)}$ is the embedding for node $i$ in layer $l$, 
$N_i$ is the set of neighbors for node $i$ in $\mathcal{G}$, and $a_l$ is the readout coefficient for each layer-$l$'s embeddings.
With the obtained graph-based user and item ID embeddings, we can construct the input features following Equation~(\ref{eq:enc_2}).
These input features can further be fed into any ranking method (e.g., DCN) to predict the interaction probabilities/ranking scores.

\subsection{The Benefit of Graphs to Ranking Methods} \label{sec:verf}
In comparison with user and item ID embeddings obtained from a table encoder as described in Section~\ref{sec:cacf}, those obtained from a graph encoder as introduced in Section~\ref{sec:graph-enc} possess additional graph topological knowledge.
To empirically verify the benefit of such graph knowledge to CTR methods, we design an experiment where all comparison models are identical except their encoders.
Specifically, we compare the ranking performance of a DCNV2~\cite{wang2021dcn} with a table encoder and that of a DCNV2 with a graph encoder.
Since the graph encoder additionally incorporates graph knowledge into ID embeddings, and these two frameworks only differ in the encoding, the ranking performance gap can indicate how integrating graph knowledge affects the ranking method (i.e., DCNV2).

In this experiment, we train models on four benchmark datasets, including Yelp2018~\cite{he2017neural}, Amazon-Books~\cite{wu2021self}, MovieLens-1M~\cite{MovieLens1M}, and Anime~\cite{AnimeRecDB}.
We evaluate the models' ranking ability with two ranking-based metrics, recall and NDCG, where larger values indicate superior ranking ability.
Their results are the averaged performance under five random seeds and are shown in Table~\ref{tab:val}, where the \textbf{Tab.} columns represent results of the DCNV2 with a table encoder, and \textbf{Graph} columns refer to those of the DCNV2 with a graph encoder.
\begin{table}
\caption{Comparative ranking performance of DCNV2~\cite{wang2021dcn} with a table encoder (Tab.), those of DCNV2 with a graph encoder (Graph), and the relative change ($\Delta\%$).}
\label{tab:val}
\centering
\begin{adjustbox}{width=\linewidth,center}
\begin{tabular}{l!{\vrule width \lightrulewidth}ccc!{\vrule width \lightrulewidth}ccc} 
\toprule
\textbf{Metric}      & \multicolumn{3}{c!{\vrule width \lightrulewidth}}{\textbf{Recall@10} $\uparrow$}              & \multicolumn{3}{c}{\textbf{Recall@20} $\uparrow$}           \\ 
\midrule
\textbf{Dataset}     & \textbf{Tab.}  & \textbf{Graph} & \textbf{\%$\Delta$}                                                       & \textbf{Tab.}  & \textbf{Graph} & \textbf{\%$\Delta$}                     \\ 
\midrule
ML-1M       & 10.73 & 11.72 & 9.25                                                      & 16.81 & 17.93 & 6.66                    \\
Yelp2018    & 4.25  & 4.30  & 1.18                                                      & 9.87  & 10.23 & 3.59                    \\
Amazon-book & 3.28  & 3.73  & 13.79                                                     & 7.53  & 8.30  & 10.23                   \\
Anime       & 15.73 & 16.97 & 7.92                                                      & 23.76 & 24.91 & 4.84                    \\ 
\midrule
\textbf{Metric}      & \multicolumn{3}{c!{\vrule width \lightrulewidth}}{\textbf{NDCG@10} $\uparrow$}                & \multicolumn{3}{c}{\textbf{NDCG@20} $\uparrow$}             \\ 
\midrule
\textbf{Dataset}     & \textbf{Tab.}  & \textbf{Graph} & \textbf{\%$\Delta$}                                                       & \textbf{Tab.}  & \textbf{Graph} & \textbf{\%$\Delta$}                    \\ 
\midrule
ML-1M       & 10.87 & 12.31 & 13.26                                                     & 12.59 & 13.93 & 10.59                   \\
Yelp2018    & 2.20  & 2.16  & -1.91                                                     & 3.75  & 3.79  & 1.23                    \\
Amazon-book & 1.92  & 2.09  & 8.98                                                      & 3.14  & 3.43  & 9.23                    \\
Anime       & 13.90 & 15.73 & 0.13                                                      & 16.41 & 18.07 & 10.09                   \\ 
\midrule
\textbf{Metric}      & \multicolumn{3}{c!{\vrule width \lightrulewidth}}{\textbf{Time (s) / Train Epoch} $\downarrow$} & \multicolumn{3}{c}{\textbf{Inference Time (s)} $\downarrow$}  \\ 
\midrule
\textbf{Dataset}     & \textbf{Tab.}  & \textbf{Graph} & \textbf{\%$\Delta$}                                                       & \textbf{Tab.}  & \textbf{Graph} & \textbf{\%$\Delta$}                     \\ 
\midrule
ML-1M       & 1.81  & 4.06  & 124.31                                                    & 0.10  & 0.24  & 140.00                  \\
Yelp2018    & 3.53  & 15.18 & 330.03                                                    & 0.17  & 0.88  & 417.65                  \\
Amazon-book & 5.56  & 29.53 & 431.12                                                    & 0.25  & 1.76  & 604.00                  \\
Anime       & 9.05  & 58.30 & 544.20                                                    & 0.47  & 3.52  & 648.94                  \\
\bottomrule
\end{tabular}
\end{adjustbox}
\end{table}
Compared with DCNV2 equipped with a table encoder, we observe that DCNV2 equipped with a graph-based encoder consistently surpasses its counterpart across the four datasets and all metrics.
The enhancement suggests that the additional graph knowledge incorporated in the graph-based ID embeddings helps improve the ranking abilities of the target ranking method.

However, this integration incurs a noticeable increase in computational overhead.
It incurs on average \textbf{$\sim$480}\% more overheads for the total training time, and \textbf{$\sim$605}\% more for the total testing time.
These excess computational overheads arise from the message passing operations described in Equation~(\ref{eq:lightgcn}) -- to acquire the ID embedding of a user/item, the model is required to query representations of all nodes within the 2-hop neighborhood of the node to conduct the further aggregation in between.
Moreover, this phenomenon can be further aggravated on dense and large graphs where the average number of neighbors per node is large.
For example, it encounters \textbf{$\sim$1049}\% more time in training and \textbf{$\sim$1261}\% more in testing in the Anime dataset.
Therefore, in industrial applications where billions of users and items construct a massive graph, simply substituting the table encoder with a graph-based encoder is prohibitively expensive and hence impractical.

\subsection{A Simple yet Effective Solution: Non-parametric Graph Convolution} \label{sec:method}
Although the introduced graph knowledge helps improve the ranking ability of the methods, significant computational overheads come along as well.
The majority of computational overheads are brought by \textbf{training with the graph encoder}, where the encoder repetitively performs the computationally expensive message-passing operation on every iteration. 

To address the acute problem of the growth of computational resources, we divert the integration of graph knowledge from the training phase to the test time, with the proposed strategy shown in Figure~\ref{fig:model} (d).
Injecting graph knowledge at testing time enjoys two benefits:
\textit{(i)} It obviates the forward passing and backpropagation entailed by message passing during training in Equation~\ref{eq:lightgcn}, whose computational overhead increases quadratically wrt the dataset density~\cite{han2022mlpinit,zhang2021graph}; 
\textit{(ii)} It avoids the computational overheads brought by repetitively performing message passing during training since it only performs message passing once at testing time. 
Our strategy can be decoupled into the following four steps:

\vspace{0.1in}
\noindent \textbf{Step 1: Constructing Similarity Matrices}

\noindent The first step is to construct two similarity matrices within users and items based on graph knowledge.
We denote the similarity matrix within users as $\textbf{A}_u$, and the matrix within items as $\textbf{A}_i$, where both depict the co-purchase relationship.
The two matrices are formulated based on the interaction matrix $\mathbf{M}$:
\begin{equation}
\begin{aligned}
    & \textbf{A}_u = \textbf{D}_u^{-\frac{1}{2}}\hat{\textbf{A}}_u \textbf{D}_u^{-\frac{1}{2}}, \hat{\textbf{A}}_u = \textbf{M}\textbf{M}^\intercal, \\
    & \textbf{A}_i = \textbf{D}_i^{-\frac{1}{2}}\hat{\textbf{A}}_i \textbf{D}_i^{-\frac{1}{2}},\hat{\textbf{A}}_i = \textbf{M}^\intercal\textbf{M}, 
\end{aligned}
\label{eq:sim}
\end{equation}
where $\textbf{D}_u \in \mathbb{R}^{|\mathcal{U}| \times |\mathcal{U}|}$ and $\textbf{D}_i \in \mathbb{R}^{|\mathcal{I}| \times |\mathcal{I}|}$ are diagonal matrices with $\textbf{D}_{u/i}[k, k]=\sum_j\textbf{A}_{u/i}[k,j]$.
Intuitively, $\hat{\textbf{A}}_{u}[i, j]$ represents the number of items interacted with both user $i$ and $j$.
Similarly, $\hat{\textbf{A}}_{i}[k, j]$ represents the number of users interacted with both item $k$ and $j$.
For popular users/items associated with massive interactions, their corresponding entries in $\hat{\textbf{A}}_{u/i}$ tend to numerically dominate the corresponding entries of unpopular users/items associated with relatively fewer interactions.
In other words, the similarity scores between popular users/items are consistently larger than those between unpopular users/items.
To remove the bias caused by node popularity (i.e., node degree), we further normalize $\hat{\textbf{A}}_{u/i}$ by accounting for the varying degree of each user/item, and denote the normalized similarity matrices as $\textbf{A}_{u/i}$.
This normalization step prevents nodes (users or items) with disproportionately high degrees from occupying overly high similarity scores~\cite{kipf2017semi}.
Therefore, we regard $\textbf{A}_{u/i}$ as the similarity matrices of users/items.

\vspace{0.1in}
\noindent \textbf{Step 2: Retrieving Relevant User/Item Candidates}

\noindent Relevant users and items based on the similarity matrices $\textbf{A}_{u/i}$ are retrieved in the second step.
Specifically, for user $i$, we obtain $n_k$ users with the corresponding top-$n_k$ similarity scores in $\textbf{A}_u[i]$, denoted as $\mathcal{U}_i$, where $\textbf{A}_u[i]$ refers to the $i$-th row of $\textbf{A}_u$.
Similarly, for item $j$, we obtain $n_k$ items with the corresponding top-$n_k$ similarity scores in $\textbf{A}_i[j]$, denoted as $\mathcal{I}_j$.
The corresponding $2n_k$ similarity scores in $\textbf{A}_{u/i}$ are extracted to further calculate the aggregation weights of the later constructed user-item pairs.
Intuitively, the higher the user/item similarity score is, the more aggregation weights should be assigned to the involved user-item pairs.

\vspace{0.1in}
\noindent \textbf{Step 3: Constructing Relevant User-item Pairs}

\begin{figure}[t]
  \centering
  \includegraphics[width=1\linewidth]{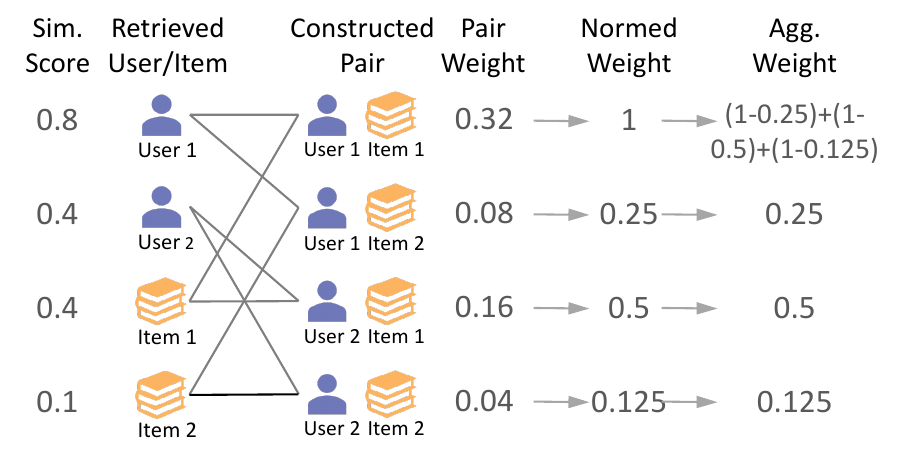}
  \caption{An example of calculating the weights for aggregating the inference results of the constructed user-item pairs. The top-2 similar users and items are retrieved to construct $2*2=4$ user-item pairs. 
  }
  \label{fig:exp}
\end{figure}

\noindent Since the users and items are selected based on the similarities in $\textbf{A}_{u/i}$, which is constructed based on the collaborative filtering signals in $\textbf{M}$, the interaction signals between the selected users and items naturally contain structural knowledge.
Consequently, a set of relevant user-item pairs is constructed by combining each user in $\mathcal{U}_i$ with each item in $\mathcal{I}_j$.
This set contains $n_k^2$ pairs and is defined as $\{\mathcal{M}_{ij}: (u, v) | \forall u \in \mathcal{U}_i, \forall v \in \mathcal{I}_j\}$.
For each user-item pair in $\mathcal{M}_{ij}$, we calculate its weight as the multiplication of the corresponding user and item similarity scores, as shown in the middle of Figure~\ref{fig:exp}.
We denote the weight matrix for all user-item pairs in $\mathcal{M}_{ij}$ as $\text{Weight}(\mathcal{M}_{ij})$, where $\text{Weight}(\mathcal{M}_{ij})[u, v]$ represents the weight for the user $u$ and item $v$ pair.

\vspace{0.1in}
\noindent \textbf{Step 4: Label Aggregation Among Relevant Ranking Scores} 

\noindent After obtaining the $n_k^2$ user-item pairs, the target ranking model is queried for the corresponding $n_k^2$ inference results, which are aggregated based on each pair's weight $\text{Weight}(\mathcal{M}_{ij})[u, v]$:
\begin{align}
    p\prime_{ij} & = \frac{\sum_{(u, v) \in \mathcal{M}_{ij}} \text{Weight}(\mathcal{M}_{ij})[u, v] * r(\textbf{z}_{uv})}{\sum_{(u, v) \in \mathcal{M}_{ij}} \text{Weight}(\mathcal{M}_{ij})[u, v]}, \\
    \textbf{z}_{uv} & = \left[ f(x_u) \; \| \; f(x_v) \; \| \; h(\textbf{c}_{uv}) \right],
    \label{eq:method}
\end{align}
where $r(\cdot)$ refers to an arbitrary trained ranking model such as the one we describe in Section~\ref{sec:cacf}. 
The aggregated inference result $p'_{ij}$ is the final re-ranking score.

Although $\text{Weight}(\mathcal{M}_{ij})$ is feasible to aggregate the inference results, we notice that the contribution proportion of the most similar user-item pair (i.e., pairs constructed with the most similar user and item) is too small.
For the example in Figure~\ref{fig:exp}, the most similar pair only contributes to $1/(1+0.25+0.5+0.125)\approx 53\%$ of the final result.
This overly small proportion may deviate the final result too much from the prediction of the original user-item pair, and thereby downgrade the performance.
To resolve this issue, we adopt a pair-wise aggregation mechanism:
\textit{(i) First}, normalize the weights in $\text{Weight}(\mathcal{M}_{ij})$ by dividing them by the maximum value in the matrix.
This step normalizes the values in $\text{Weight}(\mathcal{M}_{ij})$ between 0 to 1;
\textit{(ii) Second}, modify the aggregation weight for the most similar pair (i.e., the pair with the weight value as 1) by considering aggregating the pairwise inference result between this pair and the other pairs, as follows:
\begin{align}
    \text{Wegiht}(\mathcal{M}_{ij})[u', v'] & = \sum_{(u, v) \neq (u', v')} (1 - \text{Weight}(\mathcal{M}_{ij})[u, v]),\\ \nonumber
    \text{where}\;\; & u', v' = \argmax_{(u,v)} \text{Weight}(\mathcal{M}_{ij})[u, v].
\end{align}
The final re-ranking score is derived using Equation~\ref{eq:method} with the maximum entry in $\text{Weight}(\mathcal{M}_{ij})$ modified as above. 
In the example in Figure~\ref{fig:exp}, the proportion of the most similar pair is modified to $(1-0.25) + (1-0.5) + (1-0.125) = 2.125$. 
This modified maximum weight is approximately 71\% of the sum of the weights of all pairs, which is higher than $53\%$ before its modification.
In the following experiment section, we validate this design and empirically demonstrate how the proportion changes as $n_k$ increases.

\begin{table*}
\centering
\caption{Ranking performance comparison by applying our approach to the baselines. `\textit{Original}' and `\textit{\textbf{+Ours}}' represent performance of the original models and those applied with our strategy respectively. `\textit{\%$\Delta$}' denotes the relative performance change.}
\label{tab:main}
\begin{tabular}{l|ccc|ccc|ccc|ccc} 
\toprule

\textbf{} & \multicolumn{3}{c|}{\textbf{NDCG@10}}              & \multicolumn{3}{c|}{\textbf{NDCG@20}}              & \multicolumn{3}{c|}{\textbf{Recall@10}}            & \multicolumn{3}{c}{\textbf{Recall@20}}              \\ 
\midrule
\textbf{Model}  & \textbf{Original} & \textbf{+Ours} & \textbf{\%$\Delta$} & \textbf{Original} & \textbf{+Ours} & \textbf{\%$\Delta$} & \textbf{Original} & \textbf{+Ours} & \textbf{\%$\Delta$} & \textbf{Original} & \textbf{+Ours} & \textbf{\%$\Delta$}  \\ 
\midrule
\multicolumn{13}{c}{ML-1M}                                                                                                                                                                                                           \\ 
\midrule
\midrule
NFM             & 9.42              & 9.44         & 0.21          & 10.88             & 10.96        & 0.74          & 8.97              & 9.03         & 0.69          & 14.26             & 14.50        & 1.67           \\
DeepFM          & 10.17             & 10.41        & 2.32          & 11.70             & 12.02        & 2.72          & 9.61              & 9.98         & 3.87          & 15.08             & 15.68        & 3.98           \\
xDeepFM         & 8.08              & 8.17         & 1.06          & 9.65              & 9.82         & 1.72          & 8.04              & 8.19         & 1.92          & 13.13             & 13.52        & 3.00           \\
DCN             & 10.07             & 11.19        & 11.08          & 11.58             & 12.73        & 9.98          & 9.47              & 10.17         & 7.33          & 14.97             & 16.11        & 7.64           \\
DCNV2           & 10.76             & 10.91        & 1.39          & 12.26             & 12.52        & 2.10          & 10.02             & 10.26        & 2.39          & 15.55             & 16.12        & 3.69           \\
AutoInt         & 8.72              & 9.18         & 5.23          & 10.32             & 10.82        & 4.84          & 8.66              & 9.19         & 6.07          & 13.96             & 14.60        & 4.60           \\
EulerNet        & 8.91              & 9.11         & 2.34          & 10.51             & 10.96        & 4.32          & 8.73              & 9.24         & 5.86          & 14.08             & 14.83        & 5.28           \\ 
\hline\midrule
\multicolumn{13}{c}{Yelp2018}                                                                                                                                                                                                        \\ 
\midrule
NFM             & 2.92              & 3.35         & 14.58         & 4.58              & 5.10         & 11.35         & 5.28              & 6.06         & 14.77         & 11.23             & 12.30        & 9.58           \\
DeepFM          & 1.95              & 2.18         & 11.78         & 3.52              & 3.83         & 8.63          & 3.90              & 4.33         & 10.86         & 9.66              & 10.32        & 6.86           \\
xDeepFM         & 2.53              & 2.76         & 9.34          & 4.12              & 4.42         & 7.28          & 4.68              & 5.11         & 9.19          & 10.45             & 11.09        & 6.11           \\
DCN             & 2.20              & 2.46         & 11.73          & 3.75              & 4.12         & 10.04          & 4.25              & 4.75         & 11.78          & 9.87              & 10.75        & 8.93           \\
DCNV2           & 2.07              & 2.30         & 10.81         & 3.63              & 3.94         & 8.42          & 4.06              & 4.49         & 10.55         & 9.77              & 10.47        & 7.16           \\
AutoInt         & 1.98              & 2.20         & 11.20         & 3.53              & 3.82         & 8.33          & 3.92              & 4.32         & 10.37         & 9.57              & 10.21        & 6.64           \\
EulerNet        & 3.39              & 3.68         & 8.74          & 5.08              & 5.44         & 7.09          & 5.84              & 6.34         & 8.59          & 11.90             & 12.61        & 5.93           \\ 
\hline\midrule
\multicolumn{13}{c}{Amazon-Books}                                                                                                                                                                                                    \\ 
\midrule
NFM             & 2.67             & 3.04         & 13.69         & 4.06              & 4.47         & 10.25         & 4.52              & 5.09         & 12.42         & 9.10              & 9.81         & 7.85           \\
DeepFM          & 2.40             & 2.65         & 10.69         & 3.69              & 4.01         & 8.50          & 4.00              & 4.43         & 10.86         & 8.41              & 9.01         & 7.04           \\
xDeepFM         & 2.24             & 2.46        & 9.74          & 3.46              & 3.75         & 8.32          & 3.69              & 4.08         & 10.62         & 7.88              & 8.46         & 7.34           \\
DCN             & 1.91             & 2.20        & 15.15         & 3.13              & 3.52         & 12.40         & 3.28              & 3.81         & 16.23         & 7.53              & 8.27         & 9.94           \\
DCNV2           & 2.42             & 2.68         & 10.65         & 3.76              & 4.09         & 8.72          & 4.16              & 4.60         & 10.52         & 8.67              & 9.30         & 7.31           \\
AutoInt         & 2.35             & 2.57        & 9.63          & 3.65              & 3.94         & 7.89          & 3.93              & 4.30         & 9.47          & 8.36              & 8.92         & 6.75           \\
EulerNet        & 2.46             & 2.64        & 7.57          & 3.68              & 3.91         & 6.37          & 3.93              & 4.30         & 9.30          & 8.08              & 8.57         & 6.14           \\ 
\hline\midrule
\multicolumn{13}{c}{Anime}                                                                                                                                                                                                           \\ 
\midrule
NFM             & 14.74             & 14.68        & -0.37         & 17.60             & 17.69        & 0.55          & 17.33             & 17.71        & 2.19          & 26.11             & 26.92        & 3.10           \\
DeepFM          & 14.22             & 14.58        & 2.56          & 16.55             & 17.59        & 6.30          & 17.09             & 17.70        & 3.61          & 26.02             & 26.85        & 3.20           \\
xDeepFM         & 15.09             & 15.08        & -0.01         & 18.19             & 18.30        & 0.60          & 18.12             & 18.36        & 1.34          & 27.50             & 28.11        & 2.23           \\
DCN             & 13.90             & 14.17        & 1.94          & 16.41             & 17.03        & 3.77          & 15.73             & 16.50        & 4.90          & 23.76             & 25.26        & 6.32           \\
DCNV2           & 14.98             & 15.30        & 2.10          & 17.66             & 18.12        & 2.62          & 17.08             & 17.73        & 3.79          & 25.65             & 26.72        & 4.19           \\
AutoInt         & 12.32             & 13.27        & 7.66          & 15.00             & 16.05        & 7.04          & 14.59             & 15.99        & 9.54          & 22.75             & 24.44        & 7.45           \\
EulerNet        & 12.91             & 13.42        & 3.95          & 15.39             & 16.03        & 4.20          & 14.86             & 15.73        & 5.83          & 22.74             & 23.96        & 5.37  \\
\bottomrule
\end{tabular}
\end{table*}
\section{Experiment}

\subsection{Setup} \label{sec:exp-setup}
\subsubsection{Datasets}
We select four publicly available recommendation benchmark datasets for the experiments, including Yelp2018~\cite{he2017neural} and Amazon-Books~\cite{wu2021self} (relatively sparse), as well as MovieLens-1M~\cite{MovieLens1M} and Anime~\cite{AnimeRecDB} (relatively dense).
The statistics of the datasets are shown in Table~\ref{tab:data}.
For all datasets, we convert explicit user-to-item ratings to binary labels through thresholding.
We randomly split datasets with a ratio of 0.8/0.1/0.1 for training, validation, and testing, respectively.

\subsubsection{Baselines}
We select seven models as our baselines: NFM~\cite{he2017neural}, DeepFM~\cite{guo2017deepfm}, xDeepFM~\cite{lian2018xdeepfm}, DCN~\cite{wang2017deep}, DCNV2~\cite{wang2021dcn}, AutoInt~\cite{song2019autoint}, and EulerNet~\cite{tian2023eulernet}.
NFM~\cite{he2017neural}, DeepFM~\cite{guo2017deepfm}, and xDeepFM~\cite{lian2018xdeepfm} combine the advantages of Factorization Machines~\cite{rendle2010factorization} and deep neural networks to capture complex non-linear and high-order feature interactions.
DCN~\cite{wang2017deep, wang2021dcn} learns explicit and implicit features through a cross-network and a deep neural network, respectively.
AutoInt~\cite{song2019autoint} utilizes self-attentive neural networks to learn more effective feature interactions.
EulerNet~\cite{tian2023eulernet} learns high-order interaction features by transforming their exponential powers into linear combinations of the modulus and phase of complex features.

\subsubsection{Training}
We employ the AdamW optimizer for optimization and adopt binary cross-entropy as the loss function to train the models on the training set.
We run a fixed number of grid searches over all the baseline models' provided hyper-parameters for their best AUC performance on the validation set for the CTR prediction task.
With the best hyper-parameters, we train the models under five random seeds and save all the checkpoints.
The training and inference processes are conducted on an NVIDIA RTX 3090 GPU with 24 GB of memory, and the user-user and item-item similarity matrices are pre-computed on a standard commercial CPU with 128 GB of RAM.
We adopt Recbole~\cite{zhao2021recbole} to conduct all the experiments.

\begin{table}
\caption{The statistics of four benchmark datasets.}
\label{tab:data}
\begin{adjustbox}{width=\linewidth,center}
\centering
\begin{tabular}{lrrrc} 
\toprule
\textbf{Dataset}      & \textbf{\#User} & \textbf{\#Item} & \textbf{\#Interaction} & \textbf{Sparsity}  \\ 
\midrule
ML-1M        & 6,041    & 3,261    & 998,539         & 0.9493    \\
Yelp2018     & 77,278   & 45,639   & 2,103,896        & 0.9994    \\
Amazon-Books & 68,498   & 65,549   & 2,954,716        & 0.9993    \\
Anime        & 55,119   & 7,364    & 6,270,078        & 0.9846    \\
\bottomrule
\vspace{-0.25in}
\end{tabular}
\end{adjustbox}
\end{table}
\subsubsection{Evaluation}
For each baseline model, we evaluate our strategy with the inference results from the previously trained and saved model checkpoints.
We tune the number of candidates $n_k$ with the same amount of grid-searches and re-evaluate the model performance.
The ranking ability is evaluated by two ranking-based metrics, NDCG@K and Recall@K, both of which assign higher scores to models with stronger ranking abilities.
All reported results are averaged over the results under the five random seeds.

\subsection{Ranking Performance Improvement} \label{sec:exp-main}
We evaluate the baseline models' ranking performance on the four benchmark datasets, and denote the results under the \textbf{Original} columns in Table~\ref{tab:main}.
We apply our strategy to each of the baseline models, and report the re-evaluated ranking performance under the \textbf{+Ours} columns in Table~\ref{tab:main}.
The relative performance change under the \textbf{\%$\pmb{\Delta}$} columns is wrt the original method.

From the table, we observe that: 
\textit{(i)} our strategy stably improves the performance over the original model, and only demonstrates a slight performance downgrade in some rare cases (i.e., NFM and xDeepFM in Anime).
These results validate that our strategy is generally effective in improving ranking performance across various ranking models and benchmark datasets.
\textit{(ii)} The relative improvements are relatively evident in sparse datasets (i.e., Yelp2018, Amazon-Books) than dense datasets (i.e., ML-1M, Anime).
This is because, in sparse datasets where the number of interactions associated with each node is relatively small, the node embeddings receive fewer collaborative filtering signals from their neighbors during training.
Therefore, at test time, our strategy can compensate for insufficient training of the user/item embeddings with the injected graph knowledge.
In contrast, user/item embeddings trained in dense datasets receive more training signals from a larger number of neighbors, resulting in less space for improvement at testing time by strategy.

\begin{figure}[t]
  \centering
  \includegraphics[width=1\linewidth]{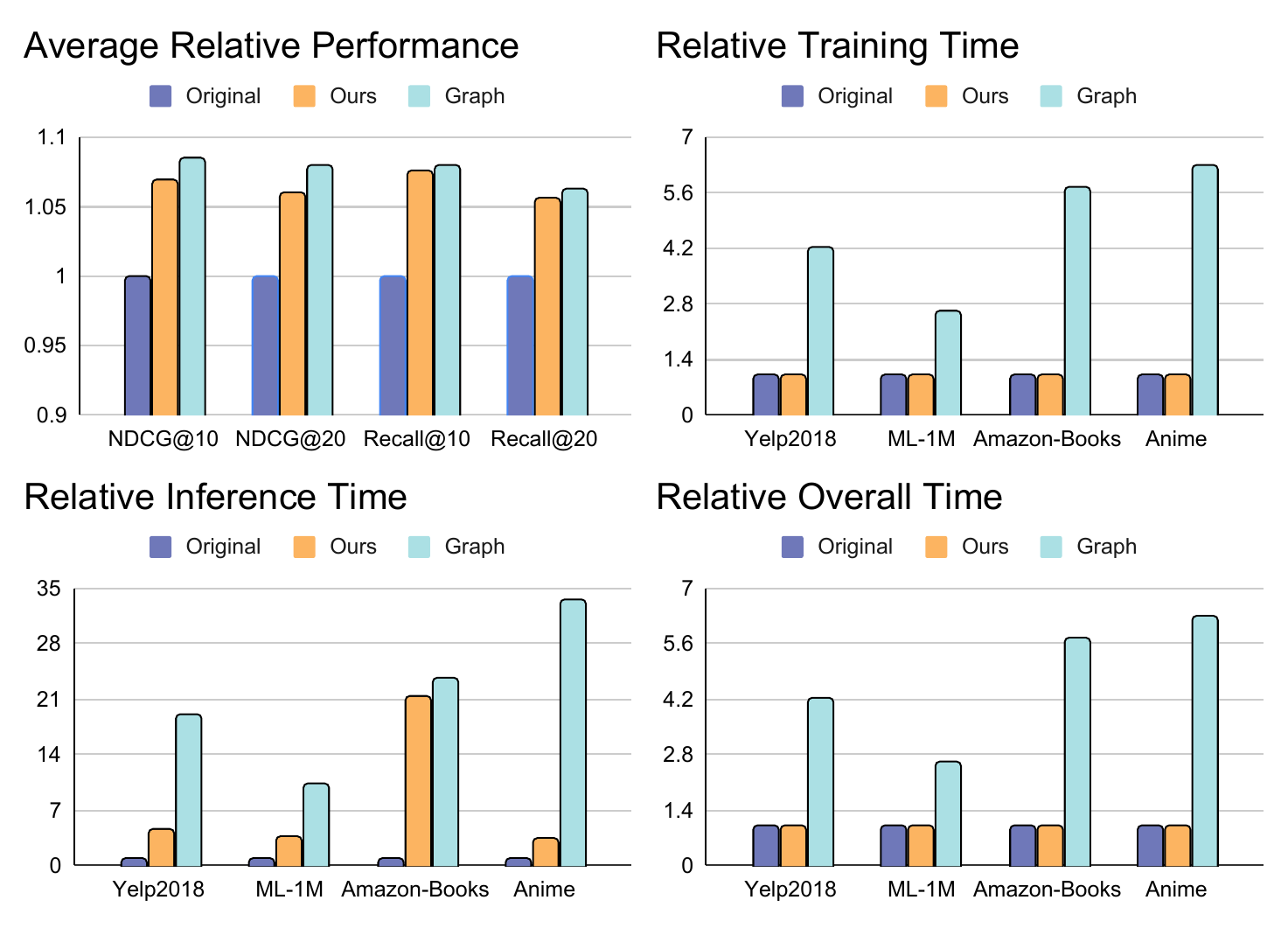}
  \caption{The relative performance is averaged over the four benchmark datasets and is scaled based on the original model performance. 
  The overall time is the summation of training and inference time.
  \textit{Original} denotes the original method, \textit{Ours} denotes methods applied with our strategy, and \textit{Graph} denotes the naive framework mentioned in Sec.~\ref{sec:graph-enc}.}
  \label{fig:time}
\end{figure}
\subsection{Time Efficiency} \label{sec:exp-time}
To demonstrate the efficiency of our strategy, we again take DCN as a typical ranking model to compare the averaged training, inference, and overall time in the three settings: DCN, DCN enhanced with our strategy, and the naive graph-enhanced DCN.
The results shown in Figure~\ref{fig:time} indicate that applying our strategy to DCN achieves a performance comparable to that of the naive graph-enhanced DCN, with only less than 2\% of extra time overall.
Applying our strategy to a well-trained CTR method does not introduce additional training time, and only quadratically increases the inference time wrt $n_k$.
For each dataset, the most relevant $n_k$ users and items neighbors can be precomputed, making the corresponding computational overhead one-off relative to a dataset.

\subsection{Effect of $n_k$ to our strategy} \label{sec:exp-nk}
\begin{figure}[t]
  \centering
  \includegraphics[width=1\linewidth]{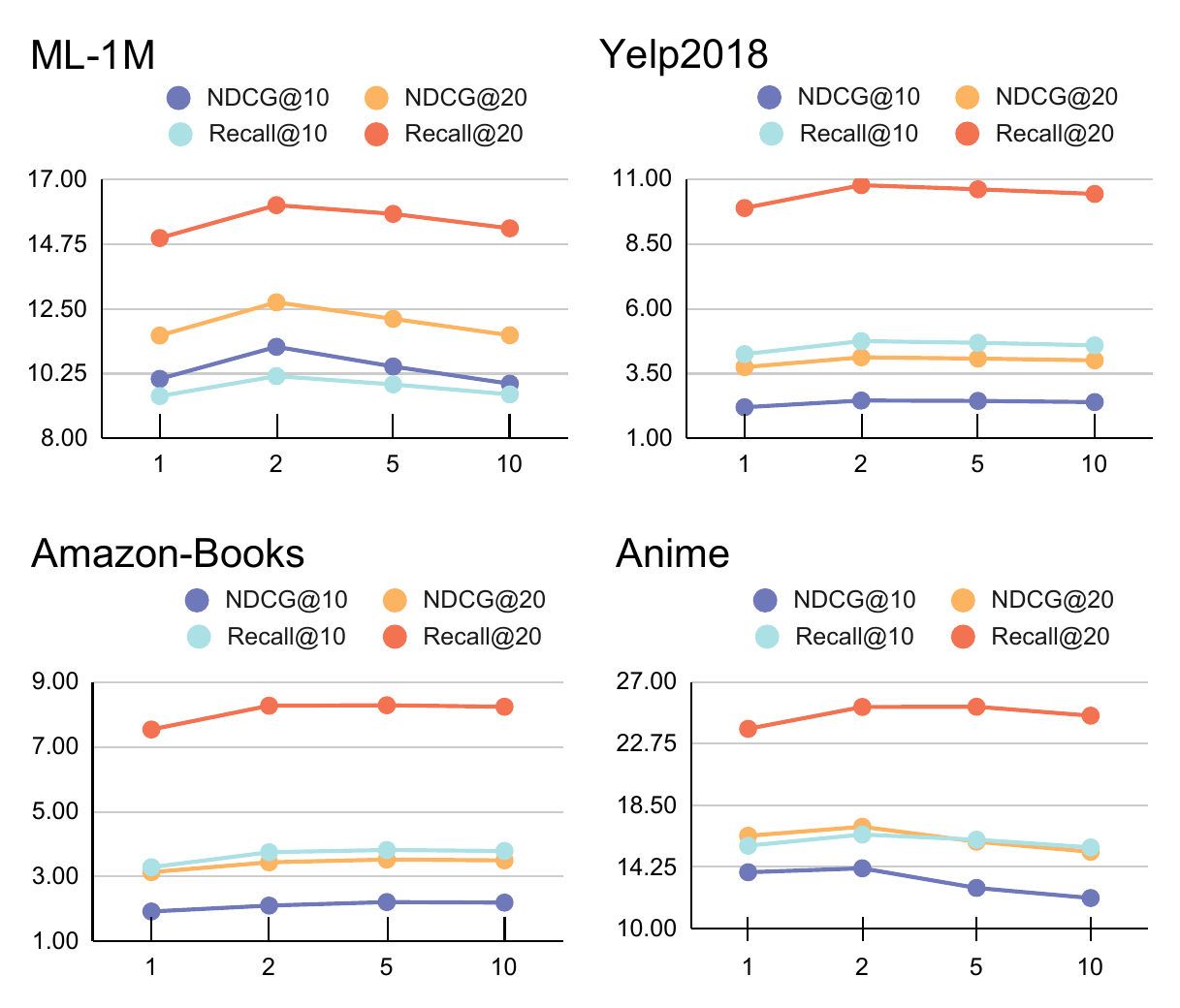}
  \caption{The ranking performance of DCN to varied number of considered neighbors $n_k$ in our strategy.}
  \label{fig:hyper}
\end{figure}

To analyze how the number of relative user/item candidates $n_k$ in our strategy affects the performance, we apply our strategy to DCN with varied $n_k$ in \{1, 2, 5, 10\} to compare the performance.
The results are shown in Figure~\ref{fig:hyper}.
From the figure, we see that our strategy, when applied to dense graphs such as ML-1M and Anime, demonstrates improved performance with smaller $n_k$'s. 
In contrast, on sparse graphs such as Yelp2018 and Amazon-books, our strategy yields better results with larger $n_k$'s.
This is because the value of $n_k$ controls the range of the neighborhood considered for graph knowledge extraction.
When $n_k$ is small, the extracted graph knowledge is sufficient to improve CTR methods trained on dense graphs but insufficient for those trained on sparse graphs.

\begin{figure}[t]
  \centering
  \includegraphics[width=1\linewidth]{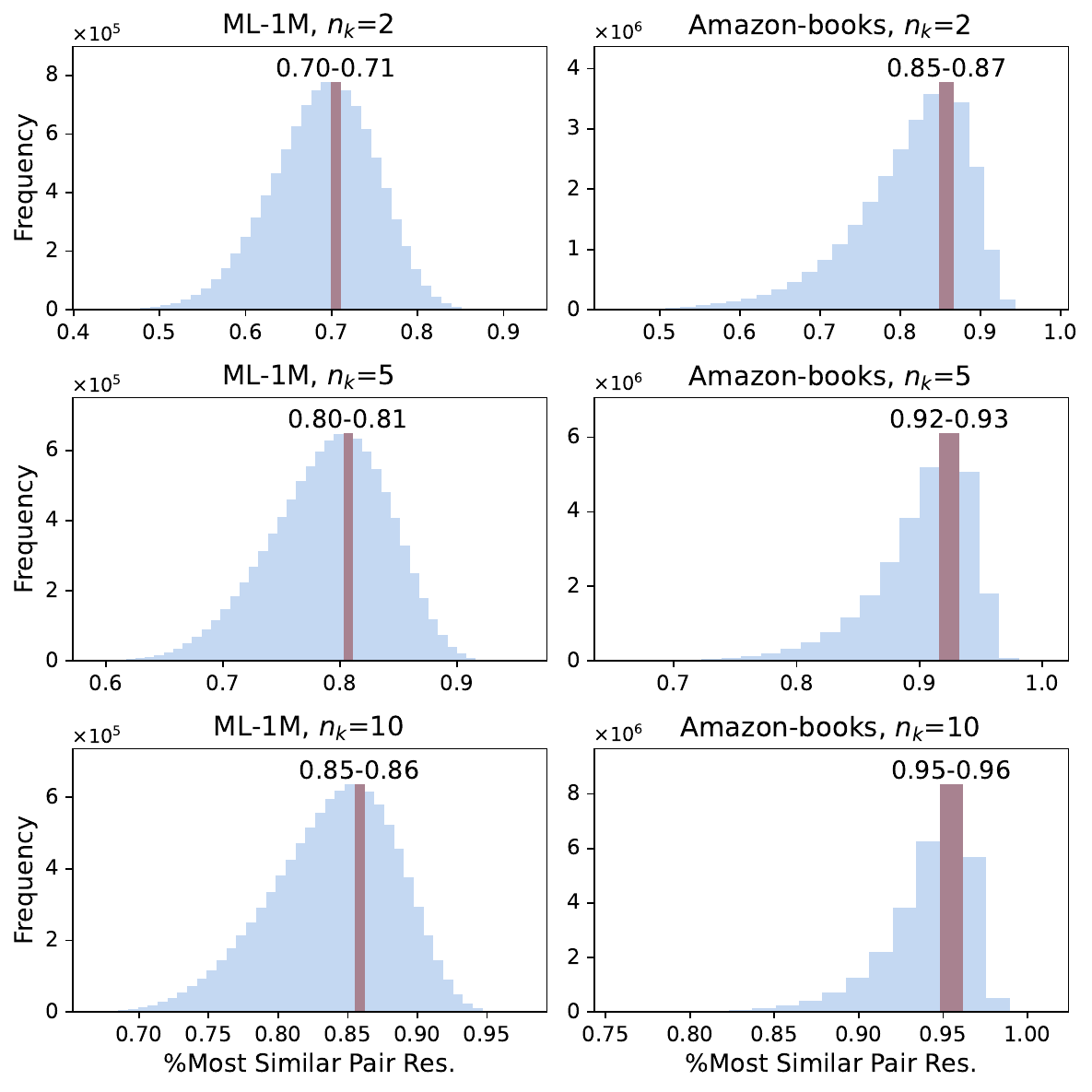}
  \caption{Contribution proportion distribution for the most similar user-item pair, where most common proportion values are highlighted and denoted above.}
  \label{fig:msp}
\end{figure}

We further analyze how $n_k$ affects the resultant contribution proportion of the most similar user-item pair in our strategy.
Specifically, we apply our strategy to DCN with varied $n_k$ in \{2, 5, 10\}, and depict the contribution proportion distribution of the most similar user-item pair in Figure~\ref{fig:msp}.
From the figure, we see that as $n_k$ increases
\textit{(i)} the contribution proportion for the most similar user-item pair increases, and
\textit{(ii)} the distribution is more concentrated (i.e., it spans fewer values).
Intuitively, the two phenomena suggest that within a close neighborhood (i.e., $n_k$ is small), our strategy adjusts the contribution proportion of the most similar pair in a wider range with relatively smaller numerical values.
Conversely, within an extensive neighborhood (i.e., $n_k$ is large), our strategy conservatively adjusts the proportions in a tighter range with relatively larger values.
\section{Related Work}
A real-world RecSys consists of roughly two stages: candidate retrieval and item ranking.
In the retrieval stage, \textit{collaborative filtering} (CF) is commonly utilized, whereas in the ranking stage where rich contextual information is additionally incorporated, \textit{click-through rate} (CTR) models are widely adopted in this stage.

\vspace{0.07in}
\noindent \textbf{Collaborative Filtering.}
As a prevalent technique that is widely employed in modern RecSys, CF makes recommendations based on the idea that similar users tend to have similar preferences~\cite{wei2023lightgt}.
Traditional CF methods
aim to reconstruct user-item interactions with parameterized user and item embeddings.
They model the reconstruction as a matrix factorization process with the user and item ID embeddings~\cite{koren2009matrix, rendle2012bpr, zhang2023lightfr}.
Some other methods maintain the ID embeddings and adopt neural networks to enhance the interaction modeling in between~\cite{he2017neural, tay2018latent}.
Apart from improving interaction modeling, other recent works focus on refining other aspects, such as the objectives and learning paradigms for performance enhancement~\cite{chen2020efficient, lee2021bootstrapping, wang2022towards, zhang2023empowering}.

\vspace{0.07in}
\noindent\textbf{Graph-based Collaborative Filtering.}
Apart from signals from direct interactions, high-order CF signals in the user-item bipartite graph are crucial for personalized recommendations as well.
These signals can be captured by the graph convolution operation in most graph neural networks (GNNs)~\cite{kipf2017semi, velivckovic2017graph, hamilton2017inductive}.
Prior efforts adopt GCN~\cite{kipf2017semi} to the user-item interaction graph~\cite{wang2019neural, berg2017graph, ying2018graph} to capture CF signals in the neighborhood.
Later on, LightGCN~\cite{he2020lightgcn} simplifies the graph convolution in GCN by preserving only linear neighborhood aggregation.
In addition to improving the structure of GNNs, recent works ~\cite{wu2021self, yu2022graph, lin2022improving, cai2022lightgcl, ma2022crosscbr} enforce contrastive learning constraints in training the models for improved performance.
For example, SGL~\cite{wu2021self} performs classical graph augmentation to the original bipartite graph to reinforce node representation learning via self-discrimination.
SimGCL~\cite{yu2022graph} refines the graph augmentation strategy in SGL with the perturbation of uniform noises and contrasts between the two perturbed graph views.
NCL~\cite{lin2022improving} explicitly incorporates potential neighbors into constructing contrastive pairs and defines a structure-contrastive objective to optimize.

\vspace{0.07in}
\noindent \textbf{Click-through Rate Prediction.}
As a widely adopted task for training models in the ranking stage, CTR prediction is defined as predicting the interaction likelihood between a user and an item given the user ID, item ID, and optional context features as input.
The incorporated contextual information includes user demographic features, item description, interaction timestamps, etc.
These features additionally consider the variability of user preferences for items across different contexts in which they interact with the system.
Early works in CTR seek efficient interactions between the interaction and the contextual information.
They preserve low-order feature interactions through the prominent Factorization Machines~\cite{rendle2010factorization} and seek high-order feature interactions through deep neural networks (DNNs)~\cite{he2017neural, guo2017deepfm, lian2018xdeepfm}.
Later works modify the layer design within a deep neural network to automatically learn bounded-degree feature interactions~\cite{wang2017deep, wang2021dcn}.
Some recent studies project the features to other predefined hyperspaces~\cite{song2019autoint, tian2023eulernet} for more efficient and complex feature interactions, and the other~\cite{ouyang2025scaled} improves the model robustness from the supervision perspective. 

Our strategy follows the established paradigm of CTR ranking methods. 
Unlike traditional graph-based CF approaches that integrate graph structure during training, our method leverages graph knowledge exclusively at inference time. 
This utilization of graph connectivity avoids the overhead of graph-based training while still benefiting from structural signals, making it lightweight and easily integrable into existing pipelines. 
Notably, it excludes the user-item interaction graph entirely from the training phase, ensuring modularity and scalability.

\section{Conclusion}
In this work, we investigate how to efficiently leverage graphs to improve the performance of models in the ranking stage of a RecSys, where rich contextual information is utilized.
We first demonstrate a naive graph-enhanced ranking framework, where graph knowledge is incorporated in the encoder of a ranking method via the message-passing operation. 
While this framework is empirically effective in improving ranking performance, the substantial computational overheads entailed by training with a graph encoder render this framework prohibitively expensive for real-world applications.
In light of this, we propose a non-parametric graph convolution strategy for ranking methods that utilizes graphs only once at test time to improve their ranking abilities.
Our strategy can be used as a plug-and-play module, and can be easily employed with various ranking methods with little to no additional computational cost.
We conduct comprehensive experiments across four benchmark datasets with various densities to demonstrate that our strategy brings noticeable ranking performance improvements (i.e., \underline{\textbf{8.1}\%} on average) during testing time with little to no additional computational overheads (i.e., \underline{\textbf{0.5}\%} on average).

\section*{Acknowledgment}
This work was partially supported by the NSF under grants IIS-2321504, IIS-2334193, IIS-2217239, CNS-2426514, and CMMI-2146076, ND-IBM Tech Ethics Lab Program, Dartmouth College, and Notre Dame Strategic Framework and Poverty Initiative Research Grants (2025). Any expressed opinions, findings, conclusions, or recommendations are those of the authors and do not necessarily reflect the views of the sponsors.
\newpage

\bibliographystyle{ACM-Reference-Format}
\bibliography{ref.bib}

\end{document}